\documentclass{amsart}

\newtheorem{theorem}{Theorem}[section]

\theoremstyle{definition}
\newtheorem{definition}[theorem]{Definition}

\theoremstyle{remark}

\numberwithin{equation}{section}

\usepackage{amsmath,amsfonts,amssymb}
\usepackage{graphicx}

\begin{document}

\title{Some Euler-type formulas for planar graphs}

\author{Armen Bagdasaryan}
\address{Russian Academy of Sciences, Institute for Control Sciences, \newline 
                                 \indent     65 Profsoyuznaya, 117997 Moscow, Russia
              }
\email{bagdasar@member.ams.org}






\begin{abstract}
The aim of this paper is to derive on the basis of the Euler's formula several analytical relations which hold for certain classes of planar graphs. 
\end{abstract}

\maketitle


\section{Introduction and Definitions}

It is known that for every connected simple planar graph there holds the Euler's characteristic $\chi$ --- a topological invariant, originally defined for polyhedra by the formula
\begin{equation}\label{euler}
\chi = V - E + F = 2,
\end{equation}
where $V$ is the number of vertices, $E$ is the number of edges, and $F$ is the number of faces in the given graph, including the exterior face. This formula corresponds to the special case $g=0$ (simple connectedness)  of the more general Poincar\'e formula for genus $g$ surfaces, in which $\chi\equiv\chi(g)=2-2g$.

In this note we derive several analytical relations, similar to Euler's formula, which hold for some classes of planar graphs that we introduce below.
A topological graph is a graph drawn in the plane such that its vertices are represented by points
and its edges are represented by arcs connecting the corresponding points such that no two arcs intersect except at a common endpoint.
The classes of graphs considered here are defined as follows.

\begin{definition}
Let $G_1(V,E)$ be a topological graph. We say that the graph $G_1$ belongs to the class $\Gamma_1$ of planar graphs if the following conditions hold
\begin{itemize}
	\item any two vertices in the graph $G_1$ are connected by at least two simple paths (chains) without a common vertex, i.e. the graph is 2-connected;
	\item all the interior faces have the same $\eta$-gonality;
	\item any vertex of the graph $v\in V$ has the degree $d_{G_1}(v) = 2,3,\dots,r$; for simplicity, below we use the symbol $d$ to denote the vertex' degree.
\end{itemize}
\end{definition}

The other two classes of graphs are derivative classes of $\Gamma_1$.

\begin{definition}
A graph $G_2$ belongs to the class $\Gamma_2$ of planar graphs if $G_2\in \Gamma_1$ and all the interior faces of $G_2$ are 4-gons and at least one interior vertex $v\in G_2$ has the degree $d(v)\neq 4$.
\end{definition}

\begin{definition}
A graph $G_3$ belongs to the class $\Gamma_3$ of planar graphs if $G_3\in \Gamma_1$ and $G_3$ contains at least one interior face with $\eta=3$.
\end{definition}

\section{Main Results}

Let $V^{ext}$ be the set of all exterior vertices of the graph, that is, those vertices which are incident to the exterior face, and let $V^{int}$ be the set of all the rest vertices -- interior ones, and
\begin{align*}
& v'\in V^{int}, \quad v''\in V^{ext}, \quad V^{int}\cup V^{ext} = V, \quad V^{int} \cap V^{ext} = \emptyset, \\
& \bigcup_{d=2}^{r}V_d^{int}=V^{int}, \quad \bigcup_{d=2}^{r}V_d^{ext}=V^{ext}, \\
& v'_d\in V_d^{int}\subseteq V^{int} \subset V, \quad v''_d\in V_d^{ext} \subseteq V^{ext} \subseteq V.
\end{align*}

Then we have
\begin{equation}\label{Vset}
|V| = |V^{int}| + |V^{ext}| = \sum_{d=2}^{r}|V_d^{int}| + \sum_{d=2}^{r}|V_d^{ext}|,
\end{equation}
where $|\cdot|$ denotes the cardinality of a set.

Constructing the simple graph of incidence ``vertices--edges'', one finds that the vertex $v_d$ forms $d/2$ edges. 
Hence, we get
\begin{equation}\label{Eset}
|E| = \frac{1}{2} \, \sum_{d=2}^{r} d \left(|V_d^{int}| + |V_d^{ext}|\right).
\end{equation}

Constructing the simple graph of incidence ``vertices--faces'', we get that the vertex $v'_d$ forms $d/\eta$ interior faces of the graph, since the vertex is incident to $d$ faces and each interior face is incident to $\eta$ vertices. Analogously, the number of interior faces formed by the vertex $v''_d$ equals to $(d-1)/\eta$.

From the above it follows that
\begin{equation}\label{Fset}
|F| - 1 = \frac{1}{\eta} \left[\sum_{d=2}^{r} r \, |V_d^{int}| + \sum_{d=2}^{r} (d-1) \, |V_d^{ext}|\right].
\end{equation}

Now substituting (\ref{Vset}), (\ref{Eset}) and (\ref{Fset}) into (\ref{euler}), we obtain the formula
\begin{equation}\label{euler1}
\begin{split}
\left[\sum_{d=2}^{r} |V_d^{int}| + \sum_{d=2}^{r} |V_d^{ext}|\right] -
\left[\frac{1}{2}\sum_{d=2}^{r} \, d \left(|V_d^{int}| \cdot |V_d^{ext}|\right)\right] + \\
\frac{1}{\eta} \left[\sum_{d=2}^{r} \, d |V_d^{int}| + \sum_{d=2}^{r} \, (d-1) |V_d^{ext}|\right] = 1.
\end{split}
\end{equation}

For planar graphs with vertices $v_d, \, (d=2,3,4)$, and $\eta$-gon faces, we get from the formula (\ref{euler1})
\begin{equation}\label{rel1}
\begin{split}
-\left(|V_4^{int}| + |V_4^{ext}| + \frac{1}{2} |V_3^{int}| + \frac{1}{2} |V_3^{ext}|\right) + \\
\frac{1}{\eta} \left(4|V_4^{int}| + 3|V_3^{int}| + 2 |V_2^{int}| + 3|V_4^{ext}| + 2|V_3^{ext}| + |V_2^{ext}|\right) = 1,
\end{split}
\end{equation}
where 
\begin{align*}
& |V| = |V_4^{int}| + |V_3^{int}| + |V_2^{int}| + |V_4^{ext}| + |V_3^{ext}| + |V_2^{ext}|, \\
& |E| = 2\left(|V_4^{int}| + |V_4^{ext}|\right) + \frac{3}{2} \left(|V_3^{int}| + |V_3^{ext}|\right) + \left(|V_2^{int}| + |V_2^{ext}|\right),
\end{align*}
and
$$
|F|-1 = \frac{1}{\eta} \left(4|V_4^{int}| + 3|V_3^{int}| + 2|V_2^{int}| + 3|V_4^{ext}| + 2|V_3^{ext}| + |V_2^{ext}|\right)
$$
or
$$
|F|-1 = |V_4^{int}| + |V_4^{ext}| + \frac{1}{2}|V_3^{int}| + \frac{1}{2}|V_3^{ext}|.
$$

From the formula (\ref{rel1}) we find
\begin{equation}\label{rel2}
\begin{split}
|V_4^{int}| (8-2\eta) + |V_4^{ext}| (6-2\eta) + |V_3^{int}| (6-\eta) + |V_3^{ext} (4-\eta) + \\
4|V_2^{int}| + 2|V_2^{ext}| = 2\eta.
\end{split}
\end{equation}

\begin{figure}[htb]
\centering
\includegraphics[scale=0.7]{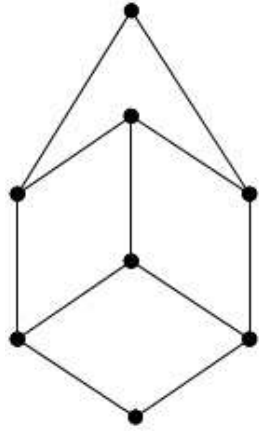}
\caption{An example of graph from the class $\Gamma_2$}
\label{fig1}
\end{figure}

Putting $\eta=4$ in the formula (\ref{rel2}), we get the relation for the class $\Gamma_2$ of planar graphs (Fig.~\ref{fig1})
\begin{equation}\label{rel3}
|V_3^{int}| + 2|V_2^{int}| = 4 + |V_4^{ext}| + |V_3^{ext}| - |V_2^{ext}|.
\end{equation}

Now assume that some given planar graph contains exactly $|F_{\eta}|$ $\eta$-gon interior faces, $|F_{\eta-1}|$ $(\eta-1)$-gon interior faces, etc., and $|F_3|$ triangular interior faces. For such planar graphs we obtain from the formula (\ref{euler1}) the following system of equations
\begin{equation}\label{systeq}
\begin{split}
 \sum_{d=2}^{r}\,d\, |V_d^{int}| + \sum_{d=2}^{r}\,(d-1)\,|V_d^{ext}| = \sum_{\eta=3}^{\eta}\,\eta \, |F_{\eta}|, \\
 -1- \left(\sum_{d=2}^{r}\,|V_d^{int}| + \sum_{d=2}^{r}\,|V_d^{ext}|\right) + 
\frac{1}{2}\,\sum_{d=2}^{r}\,d\left(|V_d^{int}| + |V_d^{ext}|\right) = \sum_{\eta=3}^{\eta}\,|F_{\eta}|.
\end{split}
\end{equation}

\begin{figure}[htb]
\centering
\includegraphics[scale=0.7]{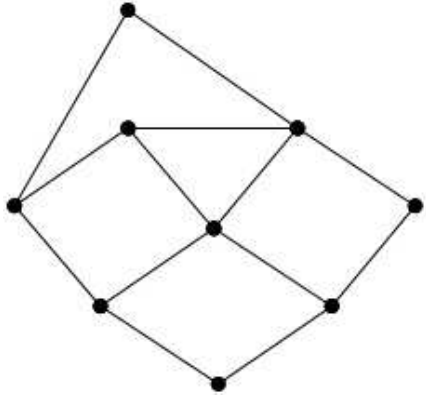}
\caption{An example of graph from the class $\Gamma_3$}
\label{fig2}
\end{figure}

From the formula (\ref{systeq}) for the class $\Gamma_3$ of planar graphs (Fig.~\ref{fig2}) we get the following relation
\begin{equation}\label{rel4}
\begin{split}
4|V_4^{int}| + 3|V_3^{int}| + 2|V_2^{int}| + 3|V_4^{ext}| + 2|V_3^{ext}| + |V_2^{ext}| = 3|F_3| + 4|F_4|, \\
1 + |V_4^{int}| + |V_4^{ext}| + \frac{1}{2}|V_3^{int}| + \frac{1}{2}|V_3^{ext}| = |F_3| + |F_4|.
\end{split}
\end{equation}

Solving the system of equations (\ref{rel4}) with respect to $|F_3|$, we obtain
\begin{equation}
4 + |V_4^{ext}| - |V_3^{int}| - 2\,|V_2^{int}| - |V_2^{ext}| = |F_3|.
\end{equation}

In the same way, using similar reasonings, one can derive analytical relations for other classes of planar graphs. 

The formulas obtained in this note can find various applications, such as deriving quantitative and qualitative estimates for algorithms of mappings of different classes of planar graphs. For instance, the relation (\ref{rel3}) allows one to recognize if the graph $G_1$ belongs to the class $\Gamma_2$ in $|V_{G_1}|$ steps, that is, by considering the $|V_{G_1}|$ rows of adjacency matrix of the graph $G_1$ vertices. In a subsequent paper we plan to deal with applications of the above formulas.

\end{document}